# AN INVESTIGATION OF MODERN FOREIGN LANGUAGE (MFL) TEACHERS AND THEIR ATTITUDES TO COMPUTER ASSISTED LANGUAGE LEARNING (CALL) AMID THE COVID-19 HEALTH PANDEMIC

Louise Hanna

School of Education, Ulster University

Hanna-l18@ulster.ac.uk

Dr. David Barr, Dr. Helen Hou, Mrs. Shauna McGill

School of Education, Ulster University

## *ABSTRACT*

*A study was performed with 33 Modern Foreign Language (MFL) teachers to afford insight into how classroom practitioners interact with Computer Assisted Language Learning (CALL) in Second Language (L2) pedagogy. A questionnaire with CALL specific statements was completed by MFL teachers who were recruited via UK based Facebook groups. Significantly, participants acknowledged a gap in practice from the expectation of CALL in the MFL classroom. Overall, respondents were shown to be interested and regular consumers of CALL who perceived its ease and importance in L2 teaching and learning.*

## 1. INTRODUCTION

The role of Computer Assisted Language Learning (CALL) has been area of interest for researchers for more than forty years (Zou & Thomas, 2019[1]). Significantly, the global Coronavirus pandemic has reinforced the importance of digital technologies in Second Language Acquisition (SLA). This study was undertaken in the summer of 2020 with 33 Modern Foreign Language (MFL) teachers in the UK as a means to comprehend their relationship with CALL at this time of a health crisis and considerable challenges in education. Overall, the investigation sought to gain an insight into MFL teachers perceive the importance, value and ease of CALL in their own teaching and learning.

Simply speaking, 'CALL refers to the application of a variety of technologies for language learning including computer, internet, online reference materials, online exercises and quizzes' (Rahimi, 2015[2]). The interdisciplinary subject of CALL has developed at breakneck speed in line with the continued evolution of digital tools and computerised technologies in education and beyond. Nonetheless, the onset of the Coronavirus outbreak heralded the most significant and radical change to the teaching and learning landscape as teachers had to adapt to the challenges of online education (Dhawan, 2020[3]). 'However, the extent to which teachers have successfully mastered these challenges and which factors are most relevant remain unknown' (König, Jäger-Biela & Glutsch, 2020[4]). Therefore, this provides the rationale to undertake this small-scale study with MFL in the UK context.

## 2. RELATED WORK

The massive global shift to online and distance learning in 2020 has been an area of significant research interest. An investigation of the attitudes of Mathematics teachers during the COVID-19 pandemic found that practitioners expressed positive opinions towards the engagement of digital devices and technological tools for the purpose of teaching and learning (Marpa, 2021[5]). In fact, a Finnish study revealed that teachers reacted quickly to learn the new technologies and perceived digital education as problematic, except for the quality of interactions with students (Niemi & Kousa, 2020[6]). With specific relation to CALL in English as a Foreign Language (EFL) instruction, teachers had 'diverse perceptions of online EFL teaching over COVID-19 as they compared it with traditional classroom language teaching to explore the features of online EFL teaching' (Gao & Zhang, 2020[7]). Furthermore, English teachers in Iran displayed positive perceptions towards the engagement of CALL for students at home whilst in lockdown (Khatoony & Nezhadmehr, 2020[8]). Overall, this study was motivated to uncover the attitudinal positions of MFL teachers towards the application of digital technologies within the UK context. In fact, 'researching teachers' beliefs are important for their professional development, particularly in the midst of a pandemic" (Zhang, 2020[9]).

## 3. METHODOLOGY

A plea for participation was issued on various MFL teaching Facebook groups. Participation entailed completing a questionnaire with specific statements relating to CALL adoption in L2 pedagogy. This type of research tool sought to obtain reactions from participants relating to their attitudes of CALL in their own teaching and learning that could be empirically measured and statistically analysed. This snapshot of teacher perceptions to the implementation of digital technologies in MFL was established, therefore, on a positivist methodology. This means that the findings are unable to account of the depth and diversity of personal and professional teacher opinion to CALL realisation. Instead, the findings provide an overview of the MFL teacher alliance with CALL that could be more thoroughly investigated via a large-scale study.

## 4. FINDINGS

Firstly, 21.1% of research participants reported having had more than twenty-years of MFL instruction. Interestingly, however, the highest level of participation involved relatively new MFL teaching practitioners who had between one to five years of MFL classroom experience (24.2%) (Table 1).

Figure 1: A table exhibiting the number of years' of MFL teaching by participants.

| Years' of teaching experience | Percentage (%) |
|---|---|
| Less than 1 year | 6.1 |
| 1-5 years | 24.2 |
| 6-10 years | 18.2 |
| 11- 15 years | 15.2 |
| 16- 20 years | 15.2 |
| +20 years | 21.1 |

In fact, these recent entrants to the teaching profession may still be transitioning from their teacher education programme to the daily demands and reality shock of the initial years of teaching. This may be particularly pronounced in the disconnect of CALL instruction in Initial Teacher Education (ITE) with the possibilities and practicalities of using digital technologies in the Newly Qualified Teacher (NQT) induction period and beyond (Woolfolk & Margetts, 2012[10]). 78.8% of study respondents agreed to the existence of such a gap between institutional expectations and implementation of CALL in the MFL classroom setting (Figure 1).

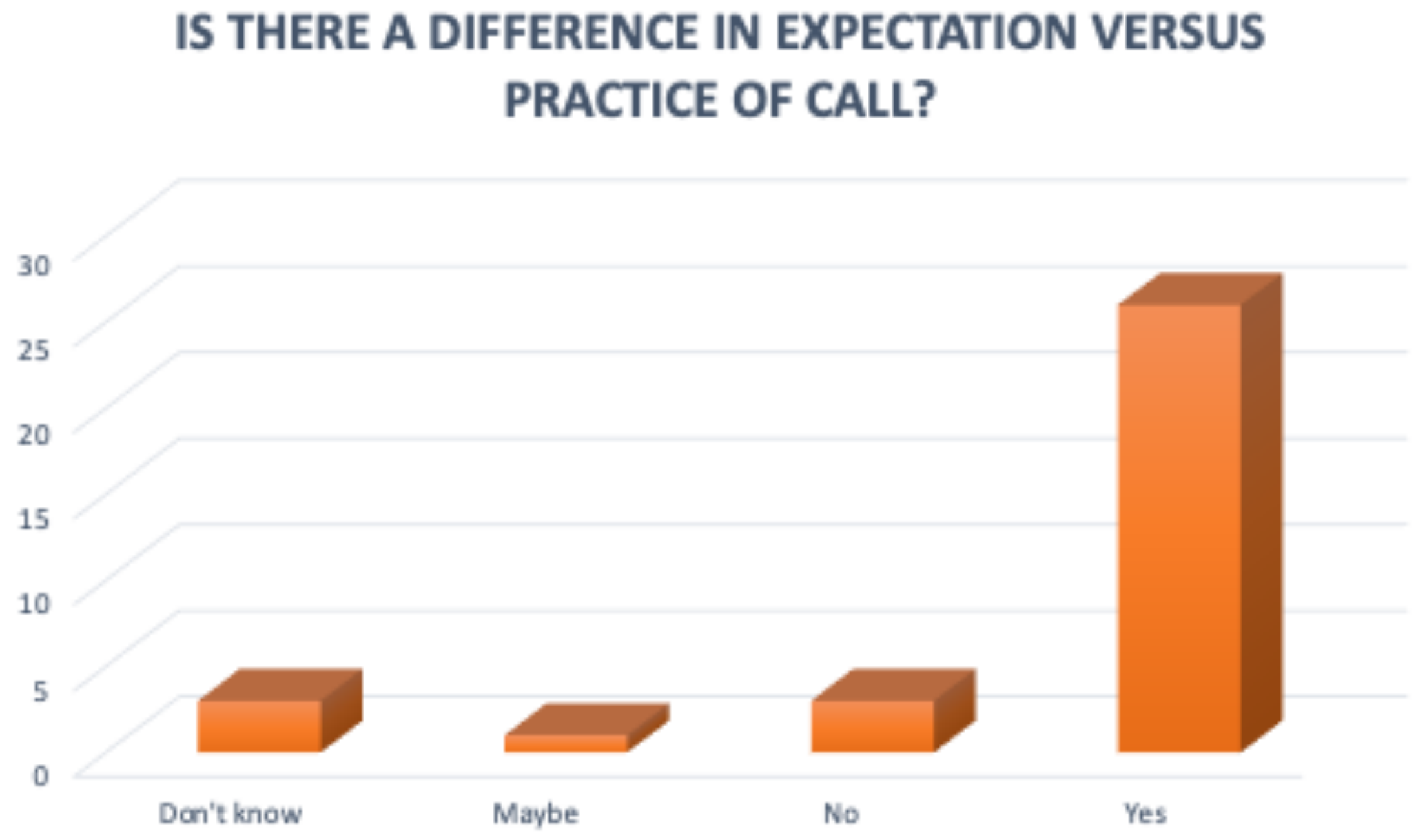

Figure 1: A bar chart highlighting the perception that there is a gap between expectation and practice of CALL usage by MFL teachers.

This technological lag has been widely documented in CALL literature as the intentions and realities of classroom innovation are mismatched (Clark-Wilson, Robutt & Sinclair, 2014[11]; Kobayashi, 2008: 105[12]). In fact, this divergence in project versus actual technological usage in pedagogy can be attributed to a number of key factors. These include concerns around the technological competencies of both teachers and learners, difficulty in accessing digital resources, issues in achieving pedagogical outcomes with CALL and a lack of comprehensive training for teaching practitioners (Visvizi, 2019[13]). Moreover, the role of educational policy can create a wedge between the expectation and the reality of CALL implementation. The small-scale study has offered a strong confirmation that a bridge between projected and actual CALL usage is desired by more than three quarters of participants (Vrasidas et al., 2006[14]).

However, this discrepancy in expectancy versus practice in CALL is not necessarily indicative of a lack of interest in digital technologies in MFL teaching and learning from the perspective of L2 teachers. This study demonstrated that 24.2% of respondents were extremely interested in CALL and 33.3% were very interested in the subject. This is presented visually in the pie chart of Figure 2. Such a claim has been authenticated - though on a higher level - in a research study by Lytras and Lytras; 70% of teachers reported an enthusiasm and readiness to adopt computer technologies and mobile innovations in their pedagogical practice (Lytras & Lytras, 2010[15]). In reality, it has often been the case that teachers have been subject to critical discourse and presented as 'outmoded, obstructive or ignorant' in relation to CALL and digital technologies (Selwyn, 2016[16]). Therefore, this could be a fruitful area of further investigation to effectively comprehend the relationship teachers have with CALL in terms of their interest to digital technologies.

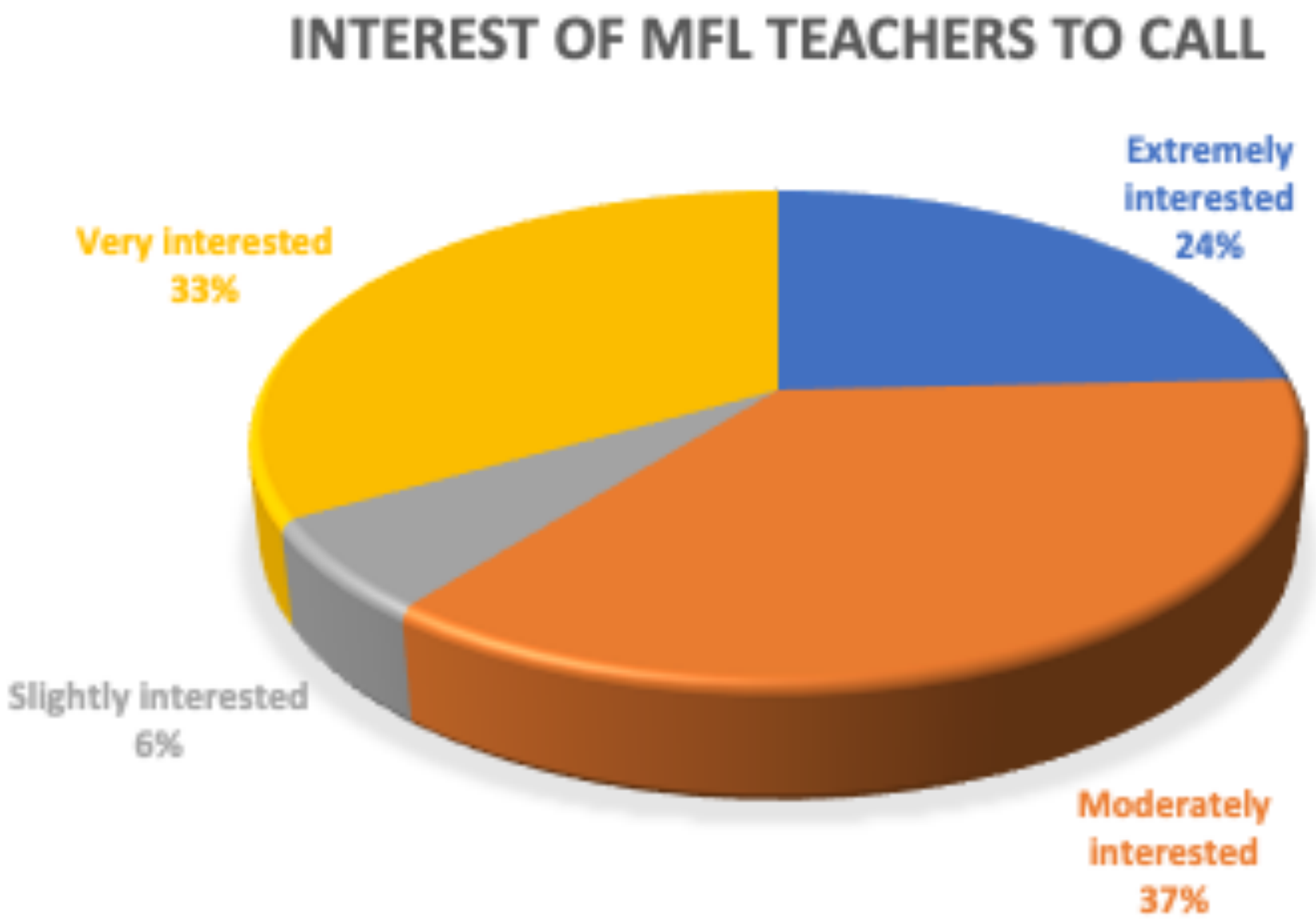

Figure 2: A pie chart showing the level of interest of MFL teacher participants to CALL.

Significantly, the finding that teachers are interested consumers of digital technologies demonstrates how the detachment between theory and practice in CALL contexts is not intrinsically bound to a lack of interest or, what is more, a perception that CALL realisation is strenuous or intellectually demanding (Lin, Zhang & Zheng, 2017[17]). In fact, 67% of respondents remarked that CALL in MFL instruction was extremely or somewhat easy (Figure 3).

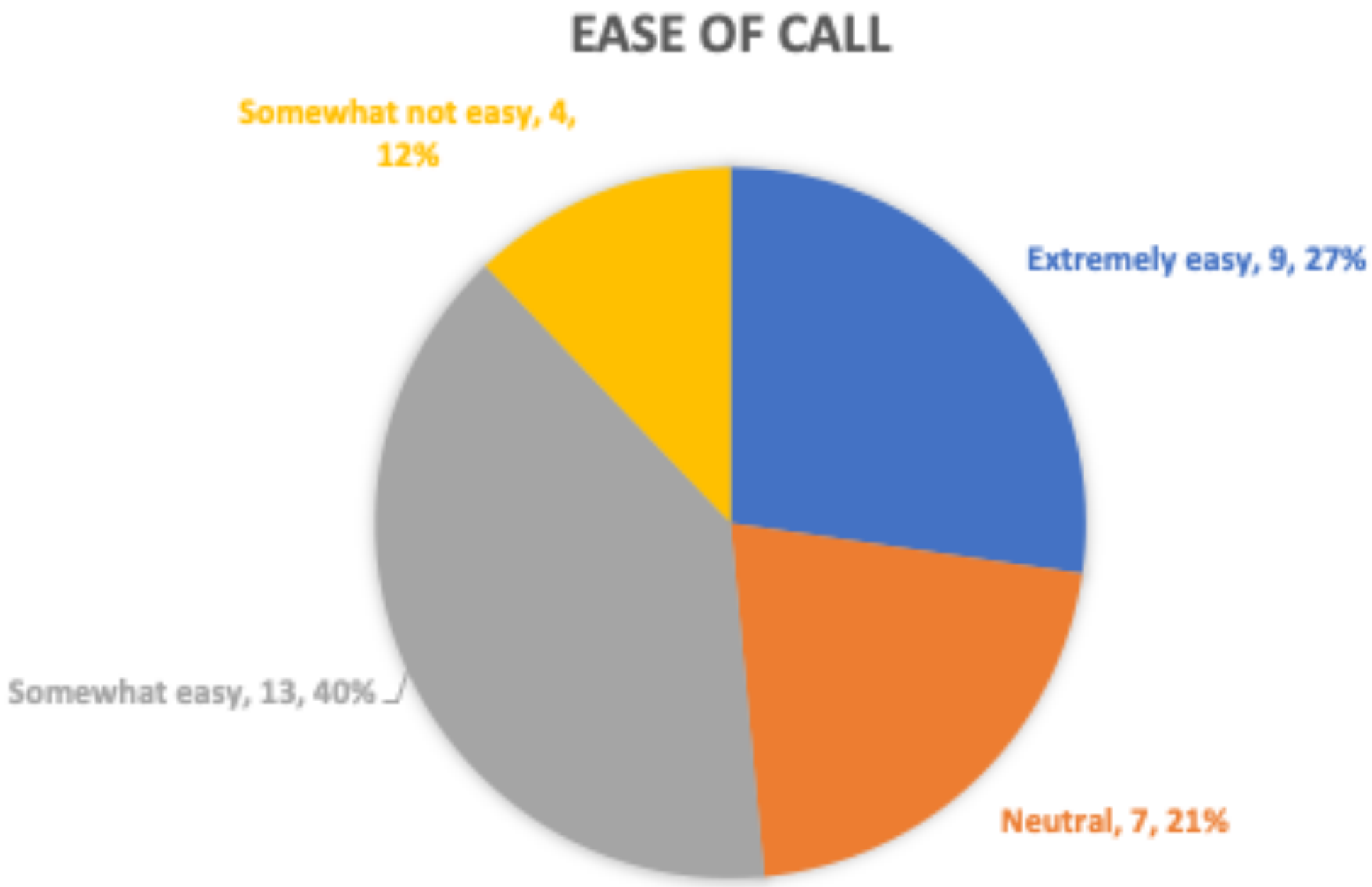

Figure 3: The perceived ease of CALL from the perception of MFL teachers.

This is an alternative perspective to the belief propagated in research literature that CALL is a complex phenomenon that constantly challenges and frequently frustrates MFL instructors (Gibson & Baek, 2009[18]; Kidd, 2008[19]). In fact, the engagement of teaching practitioners with CALL and digital technologies has been depicted as an arduous challenge (Carreira et al., 2018[20]). However, 'the products for digital teaching or learning are easier to use than the past ten years (Haghi & Luppicini, 2010[21]). In the words of one participant, adopting CALL is '*like fish in water'*. This gives the impression that innovative practice is effortless, natural and comfortable in L2 pedagogy. The results of this study point to a potentially smaller divergence in CALL practice from the perspective of the participating 33 MFL teachers. In fact, 33.3% of respondents reported always using CALL in every lesson. These findings are represented below in Table 2.

Table 2: A table depicting the frequency of CALL use in the MFL classroom.

| **Frequency of CALL usage** | **Percentage (%)** |
| --- | --- |
| Always (every lesson) | 33.3 |
| Never (not use) | 6.1 |
| Often (every other lesson) | 30.3 |
| Rarely (once a term) | 6.1 |
| Sometimes (once a month) | 24.2 |

This sample of contributing teachers are integrating digital innovations everyday into their daily MFL instruction (Bain & Weston, 2012[22]). This supports the prediction made years before that CALL would assume a normalised position in L2 pedagogy, like a pen and paper. Significantly, it was projected that CALL would be ‘used every day by language students and teachers as an integral part of every lesson’ (Torsani, 2016[23]). This regular engagement with digital technologies strongly denotes a positive belief to CALL from the research participants. This is for the reasoning that attitude to technology is inextricably linked to classroom innovation in MFL (Eshetu, 2015[24]). Overall, ‘the relationship between teacher beliefs and technology integration has also surfaced as a critical factor in technology integration’ (Brown & van der Merwe, 2015[25]). With relation to this study, participants championed the importance of CALL in MFL pedagogy. Figure 4 highlights how 37% of respondents rated CALL as extremely important and 33% appraised it as very important. This supports the claim that it is ‘important for teachers to recognize the transformative value of technology for their own practice, not just for their students’ (Pahomov, 2014[26]). This is particularly strong in the results of this study, although it is important to acknowledge the evident limitations of the investigation.

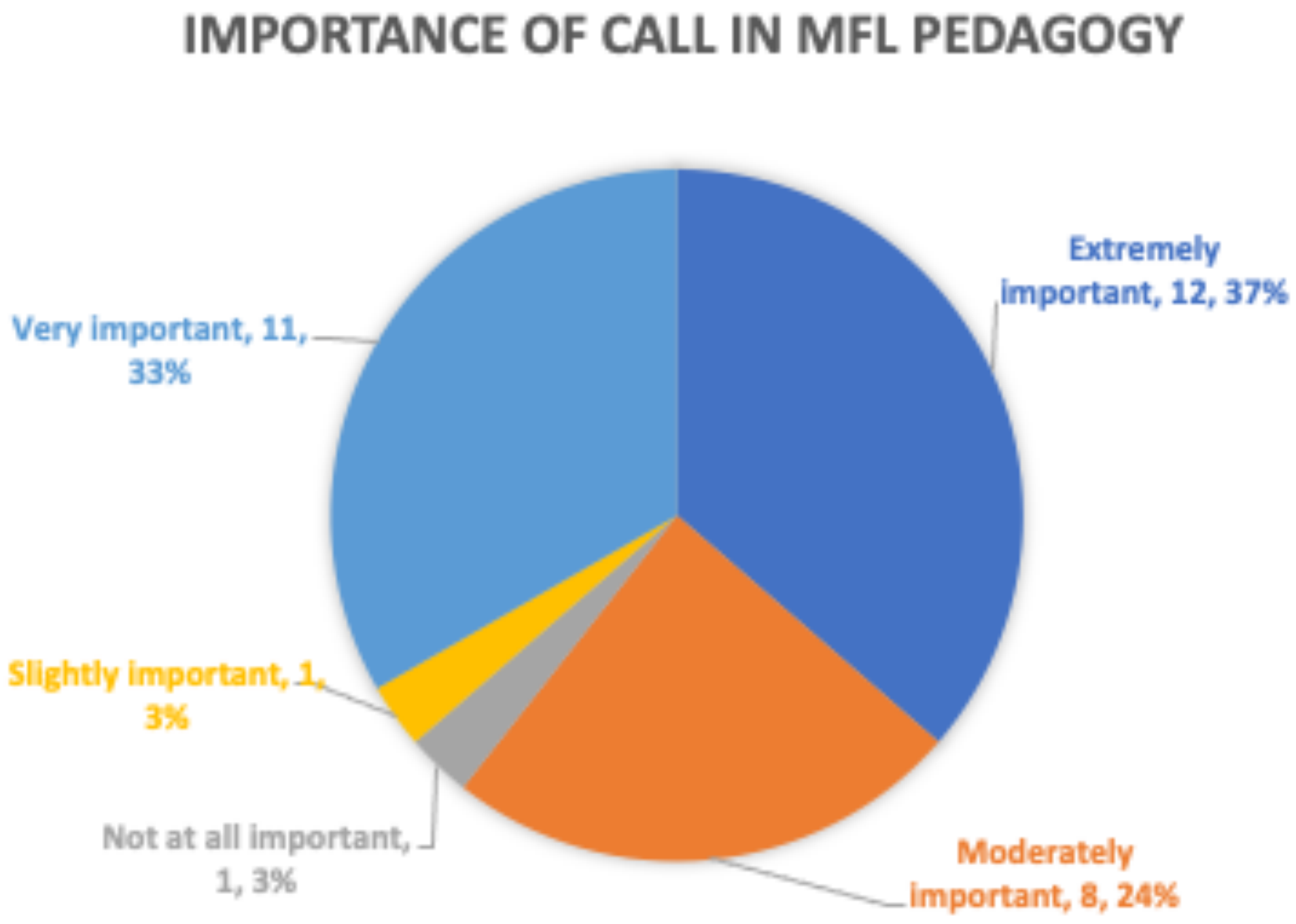

Figure 4: A pie chart presenting how important CALL is in L2 pedagogy for pilot study respondents.

Overall, MFL teachers of this study exhibited a largely positive affect to CALL which is consistent with their frequent technological practice (Ball et al., 2018[27]). A number of participants commented that CALL was ‘essential’, ‘enriching, ‘effective’, ‘exciting’, ‘helpful’, ‘invaluable’, ‘beneficial’, ‘necessary’, ‘positive’ and ‘fabulous’. However, this positive narrative to CALL was not shared by all respondents. Several teachers noted that CALL was ‘in need of direction’, ‘labour intensive’, ‘frustrating’, ‘extremely challenging’, ‘time-consuming’, ‘not the be all and end all, especially if the Internet is down’ and ‘overrated’. Nevertheless, the guidance to use CALL was offered by a third of study participants in their one piece of advice to aspiring MFL teachers (Figure 5).

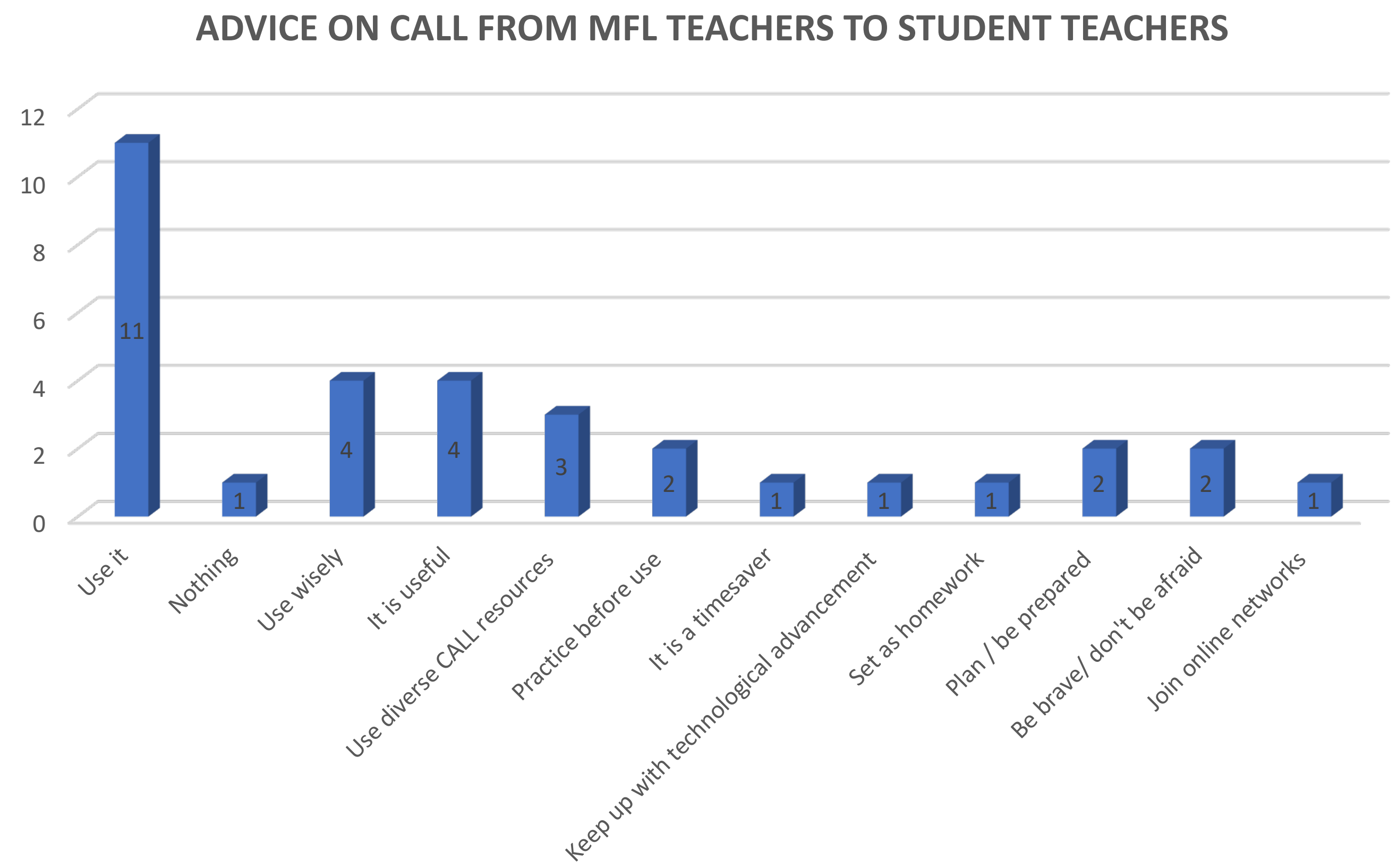

Figure 5: A bar chart representing the advice of MFL teachers to student teachers.

Such a recommendation indicates that this sample of MFL teachers were open-minded and enthusiastic in encouraging the next generation of teaching practitioners to embrace the opportunities of CALL. This has been supported in research studies investigating factors and barriers to implementation. The willingness of MFL teachers to capitalise on CALL has been thwarted externally, financially and situationally by resourcing problems, funding issues and inequitable access to infrastructure (Schul, 2019[28]; Underwood & Farrington-Flint 2015[29]). For one MFL teacher participant, CALL is only possible 'if laptops/iPads/computers room are available. I would like to be able to use more CALL as it's the way forward with technology obsessed children'. This connects to the political landscape of education and the role of Local Education Authorities (LEA) in CALL. One participant remarked that 'as decision-making stakeholders, they are the ones deciding budgets and priorities for communities'. With the onset of the COVID-19 pandemic, a number of participants remarked on the importance of General Data Protection Regulation (GDPR) and privacy in Zoom video calls. Therefore, there are wide array of factors to consider in CALL implementation for MFL teachers. Simply speaking, this study has been only able to offer a snapshot into how MFL teaching practitioners interact and relate to digital technologies in their L2 pedagogy. Therefore, the study could be viewed as a springboard from which additional investigations could be conducted to better appreciate the MFL teacher alliance to CALL.

## 5. LIMITATIONS AND FUTURE RESEARCH

A primary limitation of this study is that it was established on a positivist approach to data collection. As a consequence, conclusions are restricted by the empirical data obtained. However, richer, more detailed and in-depth information could have been acquired by a qualitative or mixed-methods approach. Such an adjustment to research methodology could have compensated for this limitation. Therefore, this opens up the possibility of conducting further investigations to better comprehend how MFL teachers perceive digital technologies. Another issue to note with this research is that it involved quite a small sample of MFL teachers (33 in total). A larger sample of participants would have enhanced the researcher's understanding of how MFL teachers interact with CALL. Future research could encompass a longitudinal understanding of MFL teachers and their relationship with computer technologies over the course of the pandemic. Additional research could be undertaken with pre-service MFL teachers to obtain a sense of their rapport with CALL while in Initial Teacher Education (ITE) during the global pandemic

## 6. Conclusions

In summary, this study presented the researcher with the occasion to gauge teacher perceptions to CALL in the L2 classroom. It has showcased that MFL teachers are daily users of CALL in L2 pedagogy. The attitudinal perspectives of participants demonstrated that a gap between expectation and practice in CALL exist -a finding that could form the basis of a follow-on study. In addition, respondents were shown to recognise the importance of digital technologies in L2 teaching and learning and readily encouraged new student teachers to adopt CALL in the classroom. In addition, participants were interested adopters of technology in the MFL classroom who perceived its usage as being easier than difficult. This, too, could be an additional study for researcher investigation. It is important to acknowledge that there are evident limitations with the study as a positivist methodology and sample size of 33 participants. Nonetheless, it has provided a snapshot of MFL teachers and their cognitions of CALL amid a global health pandemic and widespread disruption to education.

**Authors**

Louise Hanna is a second year PhD researcher at Ulster University in Northern Ireland. Her interests are extensively centred on the usage of digital technologies in L2 pedagogy. Prior to undertaking her PhD, Louise was a MFL teachers in both England and Northern Ireland.

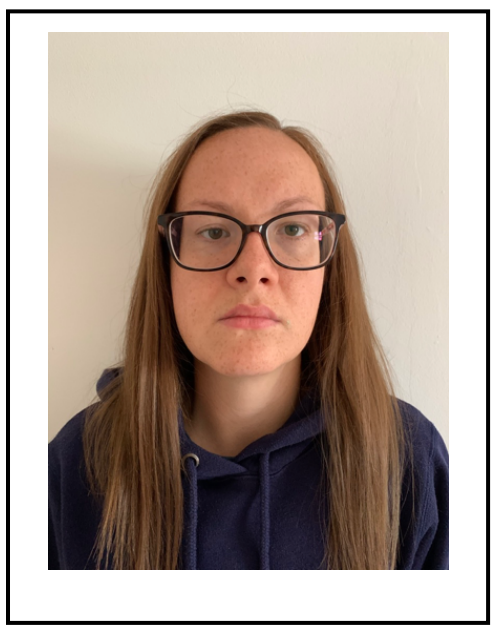